\documentclass[a4paper,11pt]{article}
\pdfoutput=1
\usepackage[utf8]{inputenc}
\usepackage{jheppub}
\setcitestyle{square,numbers}
\usepackage{amsmath}

\usepackage{extarrows}
\usepackage{bbold}
\usepackage{booktabs}
\usepackage{graphicx}
\usepackage{floatpag}
\usepackage{xcolor}
\usepackage{float}
\usepackage{multirow,array,longtable}
\usepackage{arydshln} 
\usepackage[makeroom]{cancel}
\usepackage{amsfonts}
\usepackage{stackrel}
\usepackage{slashed}
\usepackage[normalem]{ulem}

\newcommand{\be}{\begin{equation}}
\newcommand{\ee}{\end{equation}}
\newcommand{\bea}{\begin{eqnarray}}
\newcommand{\eea}{\end{eqnarray}}

\begin{document}

\begin{titlepage}

\thispagestyle{empty}

\begin{center}

\begin{Large}
\textbf{
\vglue 1.3cm
Non-relativistic neutrinos and the question of Dirac {\it vs.}\ Majorana 
neutrino nature 
} 

\end{Large}

\vspace{1cm}

{\large Evgeny~Akhmedov%
\footnote{{\tt Email:\,\href{mailto:akhmedov@mpi-hd.mpg.de}
{akhmedov@mpi-hd.mpg.de}}}%
}

\vspace{0.5truecm}

{\sl
Max-Planck-Institut f\"ur Kernphysik, Saupfercheckweg 1, 
\vspace*{0.1mm}
\\
69117 Heidelberg, Germany
}

\vspace*{12mm}

\setcounter{footnote}{0}

\textbf{Abstract}
\end{center}
\vspace*{-0.2cm}
\begin{abstract} 

Finding out if neutrinos are Dirac or Majorana particles is known to be 
extremely difficult. This is generally believed to be due to the smallness 
of the neutrino mass compared to typical neutrino energies, and to the fact 
that in the limit of vanishing mass all distinctions between Dirac and 
Majorana neutrinos disappear. This, however, does not necessarily mean that 
for non-relativistic neutrinos distinguishing between them is an easy task. 
We consider this problem in detail. The issues discussed include the 
possibilities of studying neutrino nature with non-relativistic neutrinos 
produced in $\beta$ decay in direct neutrino mass measurement experiments, 
with relic neutrinos, as well as with slow neutrinos in pair production 
processes and in scattering, both elastic and inelastic. We also discuss 
processes allowed for only one of the two neutrino types as a means of 
uncovering neutrino nature. 
\end{abstract}

\end{titlepage}


\vspace{0.2cm}

\section{Introduction}
\label{sec:Introduction}
Despite very significant experimental effort, we still do not know whether 
neutrinos are Dirac or Majorana particles. The tremendous experimental 
difficulties with finding this out are generally believed to be related to the 
extreme smallness of the neutrino mass $m_\nu$ compared to typical neutrino 
energies, as in the limit of vanishing $m_\nu$ the difference between Dirac 
and Majorana neutrinos disappears \cite{Case:1957zza,Li:1981um,Kayser:1981nw,
Kayser:1982br,Akhmedov:2024qpr}. Does this mean that for non-relativistic 
neutrinos the task would be much easer? As we shall see, not necessarily. 

Although we normally deal with highly relativistic neutrinos, there are at 
least two situations in which non-relativistic neutrinos may play a 
significant role. One is 
neutrinos of cosmic neutrino background 
(C$\nu$B). They are expected to have a nearly Fermi-Dirac spectrum with 
the present-day temperature $T_{\nu 0} = 1.945\,{\rm K} 
(=1.676\times 10^{-4}$ eV) and mean momentum $\bar{p}_\nu=3.15 T_{\nu 0}=
5.28\times 10^{-4}$\,eV.%
\footnote{We use the units $\hbar=c=k_B=1$ throughout the paper.}
{}From the values of neutrino mass squared differences measured in neutrino 
oscillation experiments it then follows that 
at least two relic neutrino species must be non-relativistic now. 
Another situation where non-relativistic neutrinos may play a role is related 
to direct neutrino mass measurement experiments, such as e.g.\ KATRIN 
\cite{Katrin:2024tvg}. These experiments aim to detect 
the electrons produced in nuclear $\beta$ decay very close to the endpoint 
of their spectrum, where a non-negligible portion of the 
emitted in the same process neutrinos should be 
non-relativistic. Non-relativistic neutrinos may also be 
encountered if there exist relatively heavy sterile neutrinos; 
in this paper we will, however, mostly concentrate on the case of 
the Standard Model with just the usual three light neutrino species 
(we will briefly comment on  sterile neutrinos in sections~\ref{sec:NC} and 
\ref{sec:1type}). We start with 
considering the production of neutrinos in charged current (CC) 
processes, such as $\beta$ decay.

\section{Non-relativistic neutrinos 
in CC processes: Direct mass measurement experiments}
\label{sec:beta}

In these experiments the spectra of electrons produced in nuclear $\beta$ 
decays, such as e.g.\  
\be
^3{\rm H}\to{\rm^3He}+e^-+\bar{\nu}_e\,,
\label{eq:tritiumBeta}
\ee 
are studied in order to gain information about the mass of the accompanying 
neutrinos. The (anti)neutrinos produced together with the electrons whose 
energies are in the immediate vicinity of the endpoint of 
their spectrum should be non-relativistic; would this 
make any observables in this process sensitive to 
neutrino nature? Would e.g.\ the electron production rate 
or their spin polarization states 
in this energy region be different for Dirac and Majorana neutrinos?%
\,\footnote{The author is grateful to Georg Raffelt for raising this   
question in an email exchange back in 2015. See also 
ref.~\cite{Millar:2018hkv}.}
Let us look into this question. 

\subsection{Beta decay: General case}

The leptonic charged current relevant to 
this process is 
\be
j^\mu_{\rm CC}(x)=\bar{\Psi}_e(x)\gamma^\mu(1-\gamma_5)\Psi_\nu(x)\,.
\label{eq:leptCC}
\ee
Here the electron quantum field operator has the plane-wave expansion 
\be
\Psi_e(x)=\int\frac{d^3p}{(2\pi)^3\sqrt{2E_{e\vec{p}}}}\sum_s \left[
b_e(p,s) u_e(p,s) e^{-i p x} + d^\dag_e(p,s) v_e(p,s) 
e^{i p x}\right], 
\label{eq:decompE}
\ee
where $b_e(p,s)$ is the annihilation operator for an electron with 4-momentum 
$p$ and in spin state $s$, and $d^\dag_e(p,s)$ is the corresponding positron   
creation operator. In the Dirac case, the neutrino quantum field 
operator can be written in a similar form:
\be
\Psi_\nu(x)=\int\frac{d^3p}{(2\pi)^3\sqrt{2E_{\nu\vec{p}}}}\sum_{s}
\left[b_\nu(p,s)u_\nu(p,s) e^{-i p x} +
d^\dag_\nu(p,s) v_\nu(p,s) e^{i p x}\right],
\label{eq:decompNuD}
\ee
with $b_\nu(p,s)$ and $d^\dag_\nu(p,s)$ being the annihilation and creation 
operators for neutrinos and antineutrinos, respectively. 

For Majorana neutrinos, the plane-wave decomposition of the quantum field is 
\be
\Psi_\nu(x)=\int\frac{d^3p}{(2\pi)^3\sqrt{2E_{\nu\vec{p}}}}\sum_{s}
\left[a_\nu(p,s)u_\nu(p,s) e^{-i p x} +
a^\dag_\nu(p,s) v_\nu(p,s) e^{i p x}\right]. 
\label{eq:decompNuM}
\ee
In this case there is no difference between neutrino and antineutrino, 
and, unlike in the Dirac neutrino case, the creation and annihilation 
operators produce and destroy particles of the same kind. 
One can readily make sure that the Majorana neutrino field operator 
(\ref{eq:decompNuM}) satisfies the self-conjugacy condition 
$\Psi^{c}_\nu(x)=\Psi_\nu(x)$. 

Consider now a $\beta^-$ decay process, first in the Dirac neutrino case.  
Let an electron and an antineutrino be produced 
with four-momenta and spins $p_1,s_1$ and $p_2, s_2$, 
respectively, The initial and the final leptonic states are then 
\be
|i\rangle=|0\rangle\,,\qquad\qquad  
|f\rangle=b^\dag_e(p_1,s_1)d^\dag_\nu(p_2,s_2)|0\rangle\,.
\label{eq:if}
\ee
Substituting the expressions for $\Psi_e(x)$ and $\Psi_\nu(x)$ from 
(\ref{eq:decompE}) and (\ref{eq:decompNuD}) into~(\ref{eq:leptCC}) and using 
the standard anticommutation rules for fermionic creation and annihilation 
operators, for the matrix element of the leptonic current one finds 
\be
\langle f|j_{\rm CC}^\mu(x)|i\rangle
=\bar{u}_e(p_1,s_1)\gamma^\mu(1-\gamma_5)v_\nu(p_2,s_2)e^{i(p_1+p_2)x}\,.
\label{eq:mElem1}
\ee

Let us now consider the Majorana neutrino case. For the final leptonic state 
 we now have, instead of the second equality of eq.~({\ref{eq:if}), 
\be
|f\rangle=b^\dag_e(p_1,s_1)a^\dag_\nu(p_2,s_2)|0\rangle\,.
\label{eq:f2}
\ee
The matrix element $\langle f|j_{\rm CC}^\mu(x)|i\rangle$ can then be  
found in the same way as it was done in the Dirac case, except that now 
eq.~(\ref{eq:decompNuM}) should be used instead of (\ref{eq:decompNuD}) 
for the neutrino field, 
and eq.~(\ref{eq:f2}) instead (\ref{eq:if}) for 
the final state $|f\rangle$. The result of this calculation turns out to 
exactly coincide with that in 
eq.~(\ref{eq:mElem1}). Thus, we conclude that 

\noindent 
\begin{itemize}
\item
{\sl All the observables in $\beta$ decay processes 
are independent of Dirac vs.\ Majorana neutrino nature}. 

\end{itemize}

\noindent 
This may appear obvious, as $\beta$ decay is an $L$-conserving process 
(in the Dirac case, when the lepton number $L$ is well defined), 
and one might think that in the Majorana case the Majorana nature of 
neutrinos may not be revealed in such processes. This is, however, incorrect: 
observables in $\Delta L=0$ processes 
in general differ for Dirac and Majorana neutrinos, the differences being 
proportional to $m_\nu$ (see e.g.\ \cite{Shrock:1982jh,Akhmedov:2024qpr}). 
 
It should be stressed that in our derivation} no assumptions have been made 
regarding the momenta of the produced neutrino states, i.e.\ the conclusion 
that $\beta$ decay is insensitive to neutrino nature is valid irrespective 
of whether the produced neutrinos are relativistic or not. Moreover, it does 
not depend on the type of the nuclear $\beta$ transition 
and on whether the parent and/or daughter nuclei 
are spin-polarized or not. Our conclusion also applies to nuclear decay by 
orbital electron capture. 

At the technical level, the absence of difference between the amplitudes of 
$\beta$ decay in the Dirac and Majorana neutrino cases 
can be understood from the following observation. 
Although, contrary to the Dirac case, Majorana neutrino 
field~(\ref{eq:decompNuM}) contains the operators $a^\dag$ and $a$  
that create and annihilate the same particles,  
only one of these two operators contributes to the matrix element 
(\ref{eq:mElem1}), depending on whether $\beta^-$ or $\beta^+$ decay 
is considered. The fact that the Majorana neutrino field 
contains the production and annihilation operators of particles of the same 
kind is therefore irrelevant.  
Obviously, this argument does not apply to processes with participation of 
more than one neutrino of the same flavor. This could be, e.g., processes 
induced by weak 
neutral currents (NC), such as neutrino scattering or pair production; 
we shall discuss them in section~\ref{sec:NC}. This could also be 
CC-induced processes of higher order in weak interaction, for instance 
double $\beta$ decay. 
  
\subsection{Detecting non-relativistic neutrinos from $\beta$ decay?}

Our conclusion that $\beta$ decay has no sensitivity to neutrino nature 
only holds true if the produced neutrinos are not observed; their 
detection could in principle allow one to discriminate between Dirac and 
Majorana neutrinos. Indeed, in the Dirac case neutrinos produced e.g. in 
$\beta^+$ decay can be detected in inverse $\beta^+$ decay reactions, but 
not in the processes of inverse $\beta^-$ decay; the latter are strictly 
forbidden by lepton number conservation. 
In contrast, in the Majorana case lepton number is not conserved, and 
detection of neutrinos produced in $\beta^+$ decays via inverse $\beta^-$ 
decay is in principle possible. However, due to the chirality selection 
rules, for relativistic neutrinos the 
capture cross section is in this case suppressed by an extremely small 
factor $\sim(m_\nu/E)^2$, which 
makes this process practically unobservable (see e.g.\ \cite{Bilenky:2006we,
Akhmedov:2024qpr}). Can non-relativistic 
neutrinos produced in $\beta$ decay experiments, for which there is no such 
suppression, be useful in this respect?  

Unfortunately, the answer seems to be negative. This is related to 
the formidable difficulties one encounters in attempting to detect 
neutrinos of very low energies, which 
up to now have prevented us from detecting the neutrinos of C$\nu$B. Actually, 
detecting the non-relativistic part of neutrinos produced in radioactive 
source experiments is much more difficult than detecting relic neutrinos, 
because the flux of the former is many orders of magnitude smaller. For 
example, the fraction of (anti)neutrinos with kinetic energies 
below 0.1\,eV produced in tritium $\beta$ decay (\ref{eq:tritiumBeta}) 
is $3.4\times 10^{-16}$; their flux at a distance as small as $10$\,cm from 
a 1\,MCi tritium source would be only $\sim 0.1\,{\rm cm^{-2} s^{-1}}$. At 
the same time, the flux of the heaviest component of C$\nu$B neutrinos at 
the Earth is expected to be
$\sim 1.7\times 10^{10}\,{\rm cm^{-2} s^{-1}}$, 
assuming no gravitational clustering.  
Clearly, as difficult as it is, it would be much easier to observe relic 
neutrinos than to distinctly detect a non-relativistic fraction of 
neutrinos produced in a radioactive source experiment (or 
coming from any other known neutrino source, including solar neutrinos and 
reactor antineutrinos).%
\footnote{
For allowed $\beta$ decays the fraction of (anti)neutrinos produced together 
with $e^\pm$ whose energies are in a narrow interval 
$\delta$ just below the endpoint $Q_\beta$ of their spectrum  
varies between $(35/16)(\delta/Q_\beta)^3$ for $Q_\beta\ll m_e$ and 
$10(\delta/Q_\beta)^3$ for $Q_\beta\gg m_e$.
} 
There is, however, an obstacle to using relic neutrinos for finding out if 
neutrinos are Dirac or Majorana particles, which is related to 
uncertainty of their flux at the Earth. 
We shall discuss this issue in section~\ref{sec:relic}. 

\section{Neutral current induced processes }
\label{sec:NC}
Let us now consider NC processes, such as neutrino scattering and neutrino 
pair production. The usual argument that it is extremely difficult to tell 
Dirac from Majorana neutrinos at relativistic neutrino energies applies to 
NC processes as well \cite{Kayser:1981nw,Kayser:1997hj,Hannestad:1997mi,
Hansen:1997sk,Zralek:1997sa,Czakon:1999ed,Akhmedov:2024qpr}; therefore, we 
shall concentrate on NC processes with non-relativistic neutrinos. 

Consider first pair production. Neutrino pairs can be produced e.g.\ in 
charged lepton annihilation processes, 
such \vspace*{-1mm}as  
\be
e^+e^-\to Z^{0*}\to 
\nu\bar{\nu}(\nu\nu)\,,
\label{eq:pairprod1}
\vspace*{0.2mm}
\ee
where $\nu\bar{\nu}(\nu\nu)$ refers to the Dirac (Majorana) case. 
However, if only the usual three light neutrino species exist, the produced 
neutrinos will always 
be highly relativistic and thus of no interest to us here. If heavy 
sterile neutrinos exist and mix with the usual ones, their pair-production in 
charged lepton annihilation processes near threshold would yield 
non-relativistic neutrino states. The production cross sections would then 
be very different for neutrinos of different nature: $\propto \beta$ for 
Dirac heavy neutrinos and $\propto \beta^3$ for Majorana ones, where 
$\beta$ is the neutrino velocity. Their angular 
distributions would also be different in the non-relativistic regime. Such 
neutrinos could in principle be detected through their subsequent decay, 
and their Dirac/Majorana nature could be established 
\cite{Ma:1989jpa,Kogo:1991ec,Hofer:1996cs,Balantekin:2018ukw}.  

Coming back to the usual light neutrinos that are of prime interest to us, 
they could also be pair-produced 
through bremsstrahlung emission, such as e.g.\ 
\be
e^{-}+X\to e^{-}+X+ \nu\bar{\nu}(\nu\nu)\,,
\label{eq:pairprod2}
\ee
and in this case kinematics allows them to be non-relativistic. However, for 
the process to be observable, it should be characterized by a noticeable 
momentum transfer, at least of keV or tens of keV range; production of 
non-relativistic pairs of light neutrinos would then correspond to a very 
small region of the allowed final-state phase space, just like in the case of 
nuclear $\beta$ decay close to the endpoint of the electron spectrum.%
\footnote{Note that the fraction of the phase space volume 
corresponding to the momenta of the two neutrinos satisfying $p<p_0$ 
is $\sim (p_0/p_{\rm max})^6$, where $p_{\rm max}$ is the maximum momentum 
allowed by the kinematics. For $p_0\sim 0.1$\,eV and $p_{\rm max}\sim 1$\,keV 
this gives $\sim 10^{-24}$.}  
The cross section of such a process would be extremely small, and therefore 
using it to uncover neutrino nature does not seem to be practical.  

Consider now NC neutrino scattering processes, a well studied 
\vspace*{-1.6mm} example being 
\be
\overset{(-)}{\nu_\mu}+e^{-}\to\overset{(-)}{\nu_\mu}+e^{-}\,.
\label{eq:nuscatt}
\vspace*{-0.6mm}
\ee
It is known that the cross sections of these processes differ for Dirac and 
Majorana neutrinos, and that the relative differences may be quite substantial 
when the involved neutrinos are non-relativistic 
(see e.g.\ \cite{Kayser:1981nw}). Elastic neutrino scattering processes have 
no physical energy thresholds, and are thus allowed for neutrinos of very 
low energies. However, for neutrino momenta $\lesssim 0.1$\,eV the kinetic 
energy of recoil electrons $T_e$ would be $\lesssim 10^{-8}$\,eV, certainly 
not a measurable quantity. On top of this, the cross section of such a 
process would be very small, \,$\lesssim 10^{-57}$\,${\rm cm^2}$. 

One might think that scattering processes in which the incoming 
neutrino is rather energetic and only the final-state one is non-relativistic 
could do the job, as the electron recoil energy could be quite substantial  
in this case. However, this would not work, because 
slow neutrinos give an advantage for studying neutrino nature in NC processes 
only when both the involved neutrinos 
are non-relativistic~\cite{Millar:2018hkv}. 
For the same reason, inelastic neutrino scattering processes, such as 
e.g.\ deuteron disintegration or nuclear excitation 
\be
\nu+d\to n+p+\nu\,,\qquad
~\nu+A(Z,N)\to A(Z,N)^*+\nu\,,~~ 
\label{eq:inelastic}
\ee
are not useful for discriminating between Dirac and Majorana neutrinos also 
in the kinematic regions where the final-state neutrinos are slow. 

Thus, NC processes with non-relativistic neutrinos do not offer a good 
alternative to other methods of studying Dirac {\em vs.}\ Majorana neutrino 
nature, such as searches for neutrinoless double $\beta$ decay or experiments 
on C$\nu$B detection, which we discuss next.

\section{Relic neutrinos}
\label{sec:relic}
\subsection{Detection problem}
\label{sec:relicDet}

The smallness of the energies of C$\nu$B  neutrinos makes their observation 
extremely difficult. A number of approaches to their detection have been 
suggested, see e.g.\ \cite{Bauer:2022lri} for a recent review. 
Unfortunately, most of them either were based on flawed considerations 
or fall short of the required sensitivity by many orders of magnitude. Out of 
all the suggestions put forward so far, only neutrino capture on beta-decaying 
nuclei \cite{Cocco:2007za,Cocco:2009rh}, such as e.g.\ 
\be
\nu_i+{\rm ^3H}\to{\rm^3He}+e^-\,,
\label{eq:tritiumInvBeta}
\ee 
has a chance to bear fruit in a foreseeable future. The approach is based 
on the separation of the spectra of $\beta$-particles produced in capture 
of relic neutrinos from those coming from the usual $\beta$-decay of the 
target nuclei. This requires energy resolution comparable to 
$m_\nu$, which is very challenging, but hopefully feasible  
\cite{Cocco:2007za,Cocco:2009rh,Betts:2013uya,PTOLEMY:2024lzs}. 

As has been mentioned in the Introduction, at least two relic neutrino 
species are currently non-relativistic. Can one use this to 
find out if neutrinos are Dirac or Majorana particles? In principle, yes; 
the reason is that the capture rate for non-relativistic relic Majorana 
neutrinos is  
twice as large as that for the Dirac ones~\cite{Long:2014zva}. 

This can be understood a follows.  Neutrinos of C$\nu$B were initially 
produced in the early Universe as relativistic particles of left handed (LH) 
and right handed (RH) chiralities, in equal amounts. 
In the Dirac case, these were LH neutrinos and RH antineutrinos;  in the 
Majorana case, these were merely neutrinos of LH and RH chiralities. 
A crucial point is that for relativistic spin-1/2 fermions chirality nearly 
coincides with helicity. Chirality is not a conserved 
quantum number, but helicity of free particles is; therefore, neutrinos 
initially produced in the LH and RH chirality states should currently be in 
the states of negative and positive helicities, respectively.   
Next observation is that, unlike in the relativistic regime, non-relativistic 
fermions of both negative and positive helicities are mixtures of LH and RH 
chiral states with nearly equal weights. Within the Standard Model, only LH 
neutrinos can be observed via inverse $\beta^-$ decay, and 
in the Majorana case non-relativistic relic neutrinos%
\footnote{Note that the method of relic neutrino detection 
based in their capture on $\beta$ decaying nuclei can only work for the 
non-relativistic components of C$\nu$B. This is because it requires 
measuring the energies of the produced electrons with energy resolution 
comparable to the mass of the detected neutrino, and achieving energy 
resolution better that $T_{\nu 0}\simeq 10^{-4}$\,eV is 
clearly impossible now (and perhaps also in the future).} 
of both helicities should contribute nearly equally to the rate of reaction 
(\ref{eq:tritiumInvBeta}) through their LH components. On the other hand, in 
the Dirac case positive helicity neutrino states are actually antineutrinos, 
as they were originally produced as states of RH chirality; their capture 
through inverse $\beta^-$ decay is thus forbidden by lepton number 
conservation. Hence, Dirac relic neutrinos can only be observed through the 
LH components of their negative helicity states. This explains the factor of 
2 difference between the detection rates of non-relativistic Dirac and 
Majorana C$\nu$B neutrinos. 
 
Can this difference be used to uncover neutrino nature once relic neutrinos 
have actually been observed? Unfortunately, there is a difficulty on this 
route. The problem is that the detection rates depend on the number densities 
and velocities of non-relativistic relic neutrino species, and their expected 
values have non-negligible 
uncertainties. While cosmology predicts the {\em average} 
densities of relic neutrinos ($n_{\nu 0}\simeq 56~{\rm cm}^{-3}$ 
per mass eigenstate and per spin degree of freedom) quite reliably, their 
densities at the Earth location may differ from that value because of 
gravitational clustering effects (see e.g. \cite{Zhang:2017ljh}).  
Both the local density and the velocities $v_i$ of the different mass 
eigenstate components of C$\nu$B should depend on neutrino mass ordering 
(normal {\em vs.}\ inverted) and on the value of the lightest neutrino mass 
\cite{Blennow:2008fh,Roulet:2018fyh}, which 
are currently unknown. It is very likely, however, that by the time 
relic neutrinos are observed, at least the neutrino mass ordering will 
have been reliably established. Still, it would be good to have a method 
of discriminating between Dirac and Majorana neutrinos that does not depend 
on the absolute value of the signal. 
 
\subsection{Unraveling neutrino nature through helicity measurements of 
relic $\nu$s?}
\label{sec:relicHelicity}
Determining Dirac {\em vs.}\ Majorana neutrino nature in relic neutrino 
detection experiments may be possible if one manages to 
determine the helicities of the observed neutrinos. Indeed, as was discussed 
above, in detection experiments based on neutrino 
capture on $\beta$ decaying nuclei non-relativistic Majorana neutrinos 
of both negative and positive helicities 
should be observed with nearly equal probabilities. 
On the other hand, in the Dirac neutrino case only 
neutrinos of one helicity can be detected (negative helicity for inverse 
$\beta^-$ decay and positive helicity for inverse $\beta^+$ decay or for 
induced orbital electron capture). Thus, detecting the helicities of the 
captured neutrinos would 
allow a determination of neutrino 
nature that does not rely on the total detection rate. 

How could one measure the helicities of relic neutrinos? This is certainly 
an extremely challenging task. One could e.g.\ get access to the information 
on neutrino helicities through their spin and velocity observations. 
For C$\nu$B neutrinos, their spin and velocity distributions might in 
principle be measured in experiments involving polarized nuclei 
\cite{Lisanti:2014pqa,Akhmedov:2019oxm}. However, 
these distributions are expected to be nearly isotropic, with only 
$\sim 10^{-3}$ anisotropy related to the peculiar motion of 
the Solar System with respect to the C$\nu$B rest frame. This means that 
a reliable detection of effects of such anisotropy 
would require unrealistically large numbers of observed 
events, \,$\gtrsim {\cal O}(10^6)$.

In principle, neutrino helicities can be measured directly, without separately 
finding the directions of their spins and momenta. The usual methods based 
on the chiral structure of weak interactions would not work for 
non-relativistic neutrinos,
as there is no one-to-one correspondence between helicity and chirality in the 
non-relativistic regime. One possible avenue would be to try to develop 
a generalization of the approach by which neutrino 
helicity was first measured in the famous experiment of Goldhaber, Grodzins 
and Sunyar (GGS) \cite{Goldhaber:1958nb}. In this experiment neutrinos were not 
detected, but their helicity was inferred, through momentum and angular 
momentum conservations, from the measurement of circular polarization of the 
de-excitation $\gamma$ rays emitted by the daughter nuclei. 

There are of course significant differences between the GGS experiment 
and relic neutrino detection via inverse $\beta$ decay. In the former case, 
nuclear decay by electron capture was used, and neutrinos were in the final 
state of the process. 
It was also crucial that the $Q$-value of the electron capture reaction was 
very close to the energy of the subsequently emitted $\gamma$ rays: this 
allowed the experimentalists to use resonance fluorescence in order to select 
$\gamma$ rays emitted in the direction of the 
recoiling nuclei without actually observing nuclear recoil. 
In C$\nu$B detection by capture on $\beta$-decaying nuclei,   
neutrinos are in the initial rather than in final state; also, they have 
very low energies, 
much smaller than the 
energies of $\gamma$ rays in any possible (and measurable) nuclear transition. 
This, in particular, means that methods based on resonance fluorescence 
cannot be employed. 

However, most of the complications related to generalization of the GGS 
method to helicity measurement of relic neutrinos can be overcome. In 
particular, daughter nuclei recoil can be measured without making use 
of resonance fluorescence. There is, however, one obstacle that will 
very likely make such a generalization impossible: the tininess of the 
momenta of relic neutrinos. Because of it, the observables in the processes 
of detection of non-relativistic relic neutrinos should be almost insensitive 
to their helicities. 
\enlargethispage{0.6cm}

\section{Processes allowed for only one neutrino type}
\label{sec:1type}

The difficulties with telling Dirac and Majorana neutrinos apart 
suggest looking for processes which, even if rare, are strictly forbidden
for neutrinos of one of the two types.%
\footnote{In this section we do not assume any restrictions on neutrino 
energies.}
Currently the most promising candidates appear to be 
neutrinoless double $\beta$-decay and related processes, 
which can only occur if neutrinos are Majorana 
particles~\cite{Schechter:1981bd} (see e.g. \cite{Dolinski:2019nrj} for a 
recent review). There are also other lepton number violating 
processes, such as e.g.\ same-sign dilepton production at colliders 
\cite{Keung:1983uu,FileviezPerez:2015mlm,Cai:2017mow}, discovery of which 
would signify Majorana nature of neutrinos.

A natural question then is:  
Are there any processes that are strictly forbidden if neutrinos are 
Majorana particles but allowed if they are of Dirac nature? 
Let us consider some candidates for such processes. 

Dirac-type neutrino mass couples LH active neutrinos with RH sterile ones. 
This could lead, for example, to active $\to$ sterile neutrino conversions  
in the early Universe, creating extra neutrino species, which could affect 
primordial nucleosynthesis 
\cite{Barbieri:1989ti,Barbieri:1990vx,Abazajian:2004aj,Abazajian:2017tcc}. 
Conversion of active neutrinos to sterile ones could also occur  
during supernova explosions, where this could drastically shorten the 
neutrino burst duration, significantly deplete supernova energy 
and even prevent  
the star from exploding \cite{Raffelt:1987yt,Gaemers:1988fp}. 
Similarly, sterile neutrino states could be produced through magnetic or 
electric dipole moments of Dirac neutrinos, either due to photon exchange in 
their scattering on electrically charged particles or through spin precession 
in external magnetic fields 
\cite{Lattimer:1988mf,Barbieri:1988nh,Gaemers:1988fp,Okun:1986na}. 
Would observation of such effects establish Dirac nature of neutrinos? 

The answer is no. If neutrinos possess Dirac mass terms, this does not 
necessarily mean that they are Dirac particles; a well known example is 
type I seesaw model, where there are both Dirac and Majorana mass terms, 
and all neutrino mass eigenstates, light and heavy alike, are Majorana 
particles. Quite similarly, the existence of Dirac-type neutrino magnetic 
moments that couple active and sterile neutrinos does not in general mean 
that neutrinos are of Dirac nature. 

Let us look at this point in more detail. Recall first that in models where 
both Dirac and Majorana masses are present, such as e.g.\ type I seesaw 
models, the general mass term in the \vspace*{-2.5mm} Lagrangian is 
\[
-{\cal L}_m=\overline{N_R}\, m_D\,\nu_L+\frac{1}{2}\overline{(\nu_L)^c}
\,m_L\,\nu_L+\frac{1}{2}\overline{N_R}\,M_R\,\overline{(N_R)^c}+h.c. 
\vspace*{-1.5mm}
\]
\be
=\frac{1}{2}
\left(\overline{(\nu_L)^c}\;\; \overline{N_R}\right)
\left(\begin{array}{cc} m_L & m_D^T\\
m_D & M_R 
\end{array}\right)
\left(\begin{array}{c}\nu_L\\ (N_R)^c
\end{array}\right)+h.c.\,,
\label{eq:massMatrix}
\vspace*{1.5mm}
\ee
where $\nu_L$ and $N_R$ are, respectively, vectors of $n$ active LH and 
$k$ sterile RH 2-component neutrino fields. 
The $(n+k)\times (n+k)$ mass matrix in the second line of 
eq.~(\ref{eq:massMatrix}) (which we denote ${\cal M}$) is symmetric, 
because so are the Majorana entries $m_L$ and $M_R$; it can be diagonalized 
through the transformation 
\be
{\cal M}={\cal U}\,{\cal M}^{diag}\,{\cal U}^T\,,
\label{eq:diag}
\ee  
where ${\cal U}$ is a unitary matrix. As is well known, in general mass 
eigenstates are Majorana particles in this case%
\footnote{This can actually be seen without explicitly diagonalizing 
${\cal M}$, by just counting the number of different mass eigenvalues and 
noting that Dirac fields are 4-component whereas Majorana ones are 
2-component.}; the exception is the case 
when both $m_L$ and $M_R$ matrices vanish. 
Similarly, most general effective Lagrangian describing the 
interaction of neutrino dipole moments with 
electromagnetic fields can be written as 
\[
{\cal L}_\mu=\Big\{\overline{N_R}\, \mu_D\,\sigma_{\mu\nu}
\nu_L+\frac{1}{2}\overline{(\nu_L)^c}
\,\mu_L\,\sigma_{\mu\nu}\nu_L+\frac{1}{2}\overline{N_R}\,\mu_R\,
\sigma_{\mu\nu}(N_R)^c\Big\}
F^{\mu\nu}
+h.c. 
\vspace*{1.5mm}
\]
\be
=\frac{1}{2}
\left(\overline{(\nu_L)^c}\;\; \overline{N_R}\right)
\left(\begin{array}{cc} \mu_L\, & -\mu_D^T\\
\mu_D\; & \mu_R 
\end{array}\right)\sigma_{\mu\nu}
\left(\begin{array}{c}\nu_L\\ (N_R)^c
\end{array}\right)
F^{\mu\nu}
+h.c.\,.
\label{eq:MagnMom}
\vspace*{1.5mm}
\ee
The real and imaginary parts of $\mu_D$, $\mu_L$ and $\mu_R$ 
are the matrices of neutrino magnetic and electric dipole moments, 
respectively. In eq.~(\ref{eq:MagnMom}) the first term in the first line 
describes transitions between active LH neutrinos $\nu_L$ and sterile RH 
$N_R$, the second term, transitions between active LH $\nu_L$ and their 
active RH CPT-conjugates, and 
the last term describes transitions between the RH and LH sterile states 
$N_R$ and $(N_R)^c=N^c_L$. CPT invariance requires the matrices of 
Majorana-type dipole moments $\mu_L$ and $\mu_R$ to be antisymmetric, whereas 
no such restriction exists for the Dirac-type matrix $\mu_D$. The matrix of 
dipole moments in the second line of~(\ref{eq:MagnMom}) is obviously 
antisymmetric; it keeps this property upon going from the flavor to mass 
eigenstates basis, because the transformations of the kind (\ref{eq:diag}) 
transform an anisymmetric matrix into an antisymmetric one. This is how it 
should be, as neutrino mass eigenstates are of Majorana nature in this case, 
and their matrix of dipole moments must be antisymmetric. Thus, the presence 
of Dirac-type dipole moments does not prevent neutrinos from being Majorana 
particles.  

Can we actually exploit the difference in the symmetry properties of 
dipole moment matrices of Dirac and Majorana neutrinos to rule out 
Majorana neutrino nature? From the antisymmetry of the matrices of dipole 
moments of Majorana neutrinos it follows that their flavor-diagonal 
magnetic and electric dipole moments 
(as well as those diagonal in the mass eigenstate basis) must vanish. 
This is because the diagonal electromagnetic moments would have to flip 
their sign upon particle-antiparticle conjugation, and 
there is no distinction between particles and antiparticles in the  
Majorana case. On the other hand, for Dirac neutrinos diagonal dipole 
moments are allowed. Let us see if this can be used 
to rule out Majorana nature of neutrinos and establish their Dirac nature. 

For definiteness, consider transitions caused by neutrino dipole moments 
in the flavor eigenstate basis. Assume that in the initial state we have 
an active LH electron neutrino $\nu_{eL}$. 
In the Majorana case, dipole moments can lead to its (spin-flavor) transitions 
to RH active neutrinos of different flavor, e.g.\ to RH active muon neutrino, 
which is usually denoted $\bar{\nu}_{\mu R}$.
However, direct transitions into RH active neutrinos of the same flavor, 
$\nu_{eL}\to \bar{\nu}_{e R}$, cannot occur. 
On the other hand, for Dirac neutrinos flavor-diagonal dipole moments are 
allowed; in particular, transitions of $\nu_{eL}$ into {\em sterile} RH states 
$\nu_{eR}$ can occur. But such a $\nu_{eL}\to \nu_{e R}$ transition is not 
the same as direct $\nu_{eL}\to \bar{\nu}_{e R}$ transition that is forbidden 
for Majorana neutrinos! In the Dirac case, we just have transitions to   
sterile states, which, as we know, are also allowed when neutrinos are 
Majorana particles. That is, here we do not have a situation when some 
process is forbidden for Majorana neutrinos but allowed for Dirac ones. 
Thus, transitions caused by electromagnetic dipole moments do not qualify 
as a means for experimentally attesting to Dirac neutrino nature.   

Actually, although direct $\nu_{eL}\leftrightarrow \bar{\nu}_{e R}$ 
processes are not allowed in the Majorana neutrino case, in the 
presence of matter they can 
still occur through a combined action of dipole moment induced spin-flavor 
transitions and the usual flavor oscillations 
\cite{Akhmedov:1991uk,Raghavan:1991em} (see e.g.\ \cite{Akhmedov:2022txm} for 
a recent discussion). Such $\Delta L=2$ processes are strictly forbidden for 
Dirac neutrinos, and thus they could serve as an alternative test for Majorana 
neutrino nature. From the existing experimental upper bounds on neutrino 
electromagnetic dipole moments it follows, however, that the probabilities of 
$\nu_{eL}\leftrightarrow \bar{\nu}_{e R}$ transitions should be very small, 
and it seems unlikely that such processes can constitute a viable competitor 
to neutrinoless double $\beta$ decay. Still, this issue 
deserves further scrutiny in our opinion. 

Interestingly, the antisymmetry of the matrix of magnetic moments of Majorana 
neutrinos leads to quadratic triangle inequalities $|\mu_{\nu_\tau}|^2\leq 
|\mu_{\nu_e}|^2+|\mu_{\nu_\mu}|^2$ and cyclic permutations, where 
$|\mu_{\nu_\tau}|^2\equiv |\mu_{\tau e}|^2+|\mu_{\tau\mu}|^2$, and the 
effective magnetic moments $\mu_{\nu_e}$ and $\mu_{\nu_\mu}$ are defined 
similarly \cite{Frere:2015pma}. Thus, experimental observation of violation of 
these inequalities could, in principle, be considered as an indication in 
favor of Dirac neutrino nature. Note, however, that this criterion is not model 
independent, as it rests on the assumption of the absence of light sterile 
neutrinos \cite{Frere:2015pma}, and their existence can never be ruled out 
experimentally.  

Summarizing, there appear to exist no processes that are allowed for 
Dirac neutrinos but forbidden for Majorana ones. It is actually not difficult 
to understand why. The cha-racteristic features of Dirac neutrinos, 
distinguishing them from Majorana ones, 
are that they are described by 4-component fields with two of the 
components being sterile, and that they carry 
conserved lepton number. However, experimentally we do not count the number 
of components of neutrino fields, and the existence of sterile neutrinos does 
not by itself mean that neutrinos are Dirac particles. 
As to lepton number conservation, processes with $\Delta L=0$ are 
allowed for both Dirac and Majorana neutrinos. 

\section{Summary and conclusions}
\label{sec:concl}

It has been quite well known that finding out if neutrinos are Dirac or 
Majorana particles is extremely difficult for relativistic neutrinos.
Our analysis shows that situation is unfortunately not much better for  
neutrinos of non-relativistic energies. For relativistic neutrinos, the 
problem is that physical observables are to a very high accuracy the same 
for neutrinos of different nature, the differences between them being 
suppressed by positive powers of $m_\nu/E$. For non-relativistic neutrinos, 
the problems are  mostly related to extremely small cross sections of their 
detection. The situation is aggravated by the fact that for most of the 
known neutrino sources the non-relativistic fractions of their fluxes are 
very small. 

Observables in direct $\beta$ decay are completely independent of 
Dirac {\em vs.}\ Majorana neutrino nature, 
irrespective of neutrino energies. Inverse $\beta$ decay processes are 
sensitive to neutrino nature in the non-relativistic domain because of 
lepton number selection rules in the Dirac case; however, detection of 
non-relativistic neutrinos through such processes meets with serious 
difficulties, which have up to now prevented us from detecting neutrinos 
of C$\nu$B. Processes induced by neutral currents have in general high 
sensitivity to neutrino nature in the non-relativistic regime, but they 
suffer from detectability problems. Namely, in elastic 
scattering processes with slow neutrinos the energies of recoiling particles 
are extremely small, which makes these processes undetectable. In neutrino 
pair production in charged lepton annihilations, the usual light neutrinos are 
produced with highly relativistic energies and therefore the observables 
exhibit the usual suppression of Dirac/Majorana differences. 
In pair production by bremsstrahlung with measurable momentum transfers 
to recoiling particles, neutrinos can only be non-relativistic in extremely 
tiny corners of the final-state phase space, and therefore the rates of such 
processes are exceedingly small. 

We have also discussed the prospects of uncovering neutrino nature through 
processes allowed for only one of the two neutrino types. The most 
propitious candidate appears to be neutrinoless double $\beta$ decay, which is 
only possible if neutrinos are Majorana particles. The experiments are, 
however, difficult, and the prospects of finding a positive signal will 
become slimmer if the current weak indications in favor of normal neutrino 
mass ordering%
\footnote{See e.g.\ \cite{Esteban:2024eli} for a recent discussion.} 
are corroborated by the future data. As an alternative, one could look for 
processes which are forbidden for Majorana neutrinos and are allowed for 
Dirac ones. Our analysis indicates, however, that such processes do not 
exist -- there is no touchstone for Dirac neutrinos.  
 
From the current perspective it appears that the best hope for studying  
Dirac {\em vs.}\ Majorana neutrino nature in the non-relativistic energy 
domain, where the differences between the two neutrino types are expected to 
be most prominent, are related to detection of neutrinos of C$\nu$B. 
This is because in experiments on relic neutrino detection by their capture 
on $\beta$ decaying nuclei the event rates for non-relativistic Majorana 
neutrinos should be twice as large as those for Dirac ones. 
The possibility of distinguishing between the two 
neutrino types thus relies on the predictions for the detection rate, which 
suffers from a number of uncertainties, most importantly, in regard to 
normal/inverted  neutrino mass ordering and to hierarchical {\em vs.}\ 
quasi-degenerate neutrino mass spectrum.%
\footnote{Note that current cosmological data 
disfavor quasi-degenerate  mass spectrum of neutrinos \cite{DESI:2024mwx}, 
though it probably would be premature to consider it ruled out. }
It is quite likely, however, that by the time relic neutrinos 
are finally detected, these uncertainties will have been cleared up.  
In any case, it is worth continuing to look for new ways of uncovering 
Dirac {\em vs.}\ Majorana neutrino nature, in particular in non-relativistic 
domain of neutrino energies. 
 
{\em Acknowledgments}.
I am grateful to Alejandro Ibarra, Gianpiero Mangano, Georg Raffelt, 
Thomas Schwetz and Alexei Smirnov for useful discussions  and to Jean-Marie 
Fr\`{e}re to drawing my attention to ref.~\cite{Frere:2015pma}.

\bibliographystyle{JHEP}
\bibliography{bibliography}

\end{document}